\documentclass{emulateapj}
\usepackage{apjfonts}
\usepackage{graphicx}

\usepackage{amsmath}

\shorttitle{Thermal Instability in First Galaxies}
\shortauthors{T. INOUE and K. OMUKAI}

\begin{document}

\title{
Thermal instability and multi-phase interstellar medium in the first galaxies
}
\author{Tsuyoshi Inoue\altaffilmark{1} and Kazuyuki Omukai\altaffilmark{2}}
\altaffiltext{1}{Division of Theoretical Astronomy, National Astronomical Observatory of Japan, Osawa 2-21-1, Mitaka, Tokyo 181-0015, Japan; tsuyoshi.inoue@nao.ac.jp}
\altaffiltext{2}{Astronomical Institute, Tohoku University, 6-3 Aramaki, Aoba, Sendai 980-8578, Japan; omukai@astr.tohoku.ac.jp}

\begin{abstract}
We examine the linear stability and nonlinear growth of the thermal instability in isobarically contracting gas with various metallicities and FUV field strengths.
When the H$_2$ cooling is suppressed by FUV fields ($G_0>10^{-3}$) or the metallicity is high enough ($Z/Z_{\rm sun}>10^{-3}$), the interstellar medium is thermally unstable in the temperature range 100-7000 K owing to the cooling by CII and OI fine-structure lines.
In this case, a bi-phasic medium with a bimodal density probability distribution is formed as a consequence of the thermal instability.
The characteristic scales of the thermal instability become smaller with increasing metallicity.
Comparisons of the nonlinear simulations with different resolution indicates that the maximum scale of the thermal instability should be resolved with more than 60 cells to follow runaway cooling driven by the thermal instability.
Under sufficiently weak FUV fields and with low metallcity, the density range of the thermal instability shrinks owing to dominance of H$_2$ cooling.
As the FUV intensity is reduced, bi-phasic structure becomes less remarkable and disappears eventually.
Our basic results suggest that, in early galaxies, i) the thermal instability has little effect for the medium with $Z/Z_{\rm sun}\lesssim 10^{-4}$, ii) fragmentation by the thermal instability could determine mass spectrum of star clusters for $10^{-4}\lesssim Z/Z_{\rm sun}\lesssim 0.04$, and iii) thermally bistable turbulent interstellar medium like our galaxy becomes ubiquitous for $Z/Z_{\rm sun}\gtrsim 0.04$, although the threshold metallicity depends on the conditions such as thermal pressure, FUV strength and redshift.
\end{abstract}


\section{Introduction}
The earliest generations of stars are believed to have played crucial roles in the cosmic evolution.
The first stars formed from the primordial pristine gas are considered to be very massive with $M\sim 100M_{\sun}$ (Omukai \& Nishi 1998, 
Bromm, Coppi, \& Larson 1999; 2002, Abel, Bryan \& Norman 2002, Omukai \& Palla 2003, Yoshida, Omukai, \& Hernquist 2008, McKee \& Tan 2008,
Hosokawa et al. 2011, Stacy, Greif, \& Bromm 2012, Hirano et al. 2014, Susa, Hasegawa, \& Tominaga 2014).
Massive stars emit a copious amount of ultraviolet (UV) photons, which ionize the surrounding interstellar medium (ISM, 
e.g., Yorke 1986, Omukai \& Inutsuka 2002, Hosokawa et al. 2011)
and eventually reionize the entire intergalactic medium (Shapiro \& Giroux 1987, Gnedin \& Ostriker 1997). 
Metal-free galaxies exposed to strong radiation fields are also considered to be favorable sites of seed supermassive-black-hole formation 
(Bromm \& Loeb 2003, Omukai, Schneider \& Haiman 2008, Dijkstra, Ferrara \& Mesinger 2014, Agarwal et al. 2014).
The massive stars inject metal-enriched ashes as well as kinetic energy into their surroundings at the end of their lives 
(Madau, Ferrara, \& Rees 2001, Mori, Ferrara, \& Madau 2002).
The accumulation of metals in the gas, raw material for star formation, elevates its cooling ability and 
thus causes a transition in the stellar mass-scale from very massive population III stars to low-mass population II stars 
(Omukai 2000, Bromm et al. 2001, Schneider et al. 2002, Omukai et al. 2005).
The kinetic energy injection controls the star formation in a complex way: the shock-compression by supernovae may promote the star formation 
(Nagakura, Hosokawa \& Omukai 2009; Kumar \& Johnson 2010), 
while evacuation of the gas from the halo would quench subsequent star formation (Bromm, Yoshida \& Hernquist 2003, Ritter et al. 2012). 
However, the physical condition of the ISM in early galaxies, which is the formation environments of second and later generation of stars, is still very uncertain, in contrast to well-defined pristine environments, where the first stars in the universe are formed.

One of the most conspicuous features of the Galactic ISM is its multi-phase nature, which is caused by thermal bistability (Field et al.  1969; Wolfire et al. 1995).
The ISM in the Galaxy is known to be thermally unstable in the temperature range between about 100 and 7000 K 
owing to the metallic fine-structure-line cooling (Field 1965).
Development of the multi-phase structure as a consequence of the thermal instability has been studied intensively in the last decade by using (magneto-)hydrodynamical simulations.
Hennebelle \& P\'erault (1999; 2000) and Koyama \& Inutsuka (2000) carried out 1D simulations of the thermally bistable ISM, and they found that 
the thermally stable warm neutral medium with number density $n\simeq1$ cm$^{-3}$ and temperature $T\simeq10^4$ K can be destabilized by shock compression, and cold clumps are produced consequently.
By using 2D simulations, Koyama \& Inutsuka (2002) showed that the thermal instability creates highly turbulent and clumpy medium in a shock compressed layer.
Detailed statistical properties of such a clumpy medium due to the thermal instability are reported by Audit \& Hennebelle (2005; 2010), Gazol et al. (2005), Gazol \& Kim (2013), Hennebelle \& Audit (2007), and Hennebelle et al. (2007), and the effect of magnetic fields on the thermal instability in multi-dimension was studied by Inoue \& Inutsuka (2008; 2009).
Heitsch et al. (2006; 2008), V\'azquez-Semadeni et al. (2006; 2007), Hennebelle et al. (2008), and Banerjee et al. (2009) studied the formation of molecular clouds from shocked warm neutral medium via thermal instability including self-gravity and magnetic field.
Recently, Inoue \& Inutsuka (2012) succeeded in reproducing the observed physical state of molecular clouds, e.g., clump mass function, turbulent velocity law etc., by carrying out numerical simulation of molecular cloud formation from atomic HI clouds 
by taking into account the chemical transitions consistently under the presence of the thermal instability and magnetic fields.
The success in those numerical studies is quite encouraging as it implies that similar numerical approach to the early-galaxy ISM is also 
viable.

Thermal instability is also operative in hot primordial gases 
with $\sim 10^4-10^6$K owing to H and He$^{+}$ bound-bound emission. 
Fall \& Rees (1985) identified clumps formed by this instability as a proto-globular cluster clouds.
In smaller halos with virial temperature $T_{\rm vir} \la 10^4$K, H$_2$ formation 
and cooling can trigger thermal instability in the temperature range of 2000-7000K 
(Sabano \& Yoshii 1977; Yoshii \& Sabano 1979, 1980; Murray \& Lin 1989; Corbelli, Galli, \& Palla 1997).
In particular, Silk (1983) pointed out that enhanced cooling by rapid three-body H$_2$ formation makes 
a collapsing pre-stellar cloud thermally unstable and triggers its fragmentation around $10^{8-9} {\rm cm^{-3}}$.
Although this instability appears too weak to cause fragmentation in reality 
(Abel, Bryan \& Norman 2002; Omukai \& Yoshii 2003; see also Greif, Springel, \& Bromm 2013), 
it might be relevant to the origin of the two-phase gas around the first protostars found in numerical simulations
(Turk, Norman \& Abel 2010).  
In a population II star-forming ISM with some metal enrichment, thermal instability is expected to be driven by 
the metallic fine-structure-line cooling as in the Galactic ISM. 
For example, Suchkov, Suchkov \& Shchekinov (1981) studied the thermal equilibrium state for the gas with metallicity 
$10^{-2}Z_{\odot}$ and discussed the fragmentation of proto-globular cluster clouds by thermal instability. 

So far, however, development of the thermal instability in the low-metallicity medium has never been confirmed
by multi-dimensional hydrodynamical simulations and its physical condition has not been studied comprehensively.  
In this paper, as the first attempt of this sort, 
we investigate the non-linear growth of thermal stability in the ISM with metallicity $Z=10^{-4}-1Z_{\sun}$ 
by way of local three-dimensional simulations.
Similar to the case of the Galactic ISM, post-shock regions are likely environments of 
thermal instability in early galaxies. 
Shocks are considered to be prevalent in forming galaxies as a result of more frequent merging with other galaxies, 
inflow of the intergalactic gas, and compression by the supernova explosion after vigorous starbursts. 
Those post-shock layers are known to contract isobarically until the postshock gas reaches thermal equilibrium (e.g., Yamada \& Nishi 1998, Koyama \& Inutsuka 2000).
Even without shocks, with metallicity $Z\gtrsim 10^{-4}\,Z_{\sun}$, 
rapid cooling causes the protogalactic clouds to contract isobarically (Safranek-Shrader et al. 2014).
Thus, here, we study thermal stability in isobarically contracting low-metallicity media.

The organization of this paper is as follows:
In \S 2, we first examine when the condition for the thermal instability is satisfied in low-metallicity ISM and 
estimate the unstable scales for various metallicities and background ultraviolet field strengths.
In \S 3, we describe the local three-dimensional simulation of the thermal-instability development in low-metallicity medium.
Finally in \S4, we summarize our findings and discuss their implications for low-metallicity star formation.

\section{Linear stability analysis}\label{s1}
Before the full simulations, we explore the density/temperature ranges of thermal instability by 
applying the instability criterion derived by the linear perturbation theory. 
Here, we take an isobarically contracting gas as the background medium against which we examine the thermal instability. 

\subsection{Isobarically contracting background}\label{s11}
The evolution of the unperturbed background state is described by the energy equation 
\begin{equation}
\frac{de}{dt}=-p \frac{d}{dt}\left(\frac{1}{\rho}\right) - \frac{\Lambda}{\rho}
\end{equation}
supplemented with the equation of state:
\begin{equation}
e=\frac{1}{\gamma-1}\frac{p}{\rho},
\end{equation}
where $e$ is the specific internal energy, $\rho$ is the mass density, $p$ is the thermal pressure, $\gamma$ is the adiabatic index, and $\Lambda$ is net cooling rate per unit volume.
Using the isobaricity $p=const.$, the above equations can be combined into the following single equation:
\begin{equation}
\frac{d\rho}{dt}=\frac{\gamma-1}{\gamma}\frac{\rho}{p}\Lambda.
\label{eq1}
\end{equation}
Here, we note again that the isobaric condition assumed for the background state is expected to hold both in the postshock cooling layer and 
the protogalactic clouds, i.e., the site of star formation in the first galaxy, until the gas reaches thermal equilibrium state, 
as long as the cooling timescale is shorter than the compression timescale by external force such as self-gravity.
In \S 2.3, we show that the isobaricity would break down in the media with $Z\lesssim 10^{-4} Z_{\odot}$.
We include a requisite minimum set of chemical reactions and cooling/heating processes, which are summarized in Tables 1 and 2, respectively.
The whole chemical network and resulting cooling rate are solved self-consistently with the background isobaric contraction.
We consider an optically thin medium, although our code is able to handle opaque cases as well (Inoue \& Inutsuka 2012).

The background far UV (FUV) intensity is expressed by so-called the Habing parameter $G_0$, which indicates 
the field strength relative to that in the solar neighborhood and is defined by the radiation energy density 
in the energy range between 6 eV and 13.6eV normalized by $5.29\times 10^{-14}{\rm erg~cm^{-3}}$ (Habing 1968).
The current most reliable estimate is by Mathis, Mezger \& Panagia (1983), who updated the solar neibhborhood value to $G_0=1.14$ (Draine 2011). 
The Habing parameter is related to another frequently-used FUV parameter $J_{21}$, which is the specific intensity at the Lyman limit 
normalized by $10^{-21} {\rm erg~s^{-1}~cm^{-2}~Hz^{-1}~str^{-1}}$:
\begin{equation} 
J_{21}=20.9~G_0 
\end{equation}
for the FUV spectral shape of Mathis, Mezger \& Panagia (1983).

We start the numerical integration of eq.~(\ref{eq1}) from a number density $n_{0}\equiv \sum n_i=10$ cm$^{-3}$ with abundances of $x_{\rm H}(0)=0.92$, $x_{\rm He}(0)=0.08$, $x_{\rm p}(0)=3\times10^{-4}\,x_{\rm H}(0)$, $x_{{\rm He}^+}(0)=3\times10^{-4}\,x_{\rm He}(0),$ and $x_{{\rm C}^+}(0)=1.4\times10^{-4}\,(Z/Z_{\odot})$, where $x_{\rm i}\equiv n_{\rm i}/n_0$.
Other abundances ($x_{{\rm H}^-}$, $x_{{\rm H}_2}$, $x_{{\rm C}}$, and $x_{{\rm CO}}$) are chosen to be zero initially. 
In the following, we indicate the metallicity $Z$ by using the parameter $[Z]\equiv {\rm log} (Z/Z_{\odot})$, 
where $[Z]=0$ indicates the solar metallicity. 
The abundances of metals, dust and polycyclic aromatic hydrocarbons (PAHs) 
are set to be proportional to the metallicity $Z$.
The fiducial value of the thermal pressure is chosen to be $p/k_{\rm B}=10^5$ K cm$^{-3}$, 
consistent with the value for the collapsing cloud with $[Z]=-3$ 
in Safranek-Shrader et al. (2014)'s simulation.

\begin{deluxetable*}{lll} \label{t1}
\tablewidth{0pt}
\tablecaption{Chemical Reactions}
\tablehead{Reaction &  Note  & Reference }
\startdata
H$^-$ photo-dissociation & depends on FUV strength $G_{0}$ (visual extinction Av=0 is assumed) & 1 \\
H$^-$ formation & collision with e & 2 \\
H$_{2}$ formation & H$^-$ and H collision & 3 \\
H$_{2}$ formation on grains & dust temperature $T_{\rm d}=10$ K is assumed & 4 \\
H$_{2}$ photo-dissociation & depends on $G_{0}$ (column density $N_{\rm{H}_2}$=0 and Av=0 is assumed)  & 5 \\
H$_{2}$ and CO dissociation & collisions with e, p, and H  & 6 \\
H$_{2}$ and CO dissociation & recombination of He$^+$  & 1 \\
H, He, and C ionization & collisions with e, p, H, and H$_{2}$ & 1,\,6\\
H$^{+}$, He$^{+}$, and C$^{+}$ recombination & --- & 6,\,7 \\
CO formation & depends on $G_0$ (Av=0 is assumed) & 8 \\
CO photo-dissociation  & depends on $G_0$ ($N_{\rm{H}_2}=N_{\rm CO}=$ Av=0 is assumed) & 8,\,9 \\
C photo-ionization & depends on $G_0$ ($N_{\rm C}=N_{\rm{H}_2}=$ Av=0 is assumed) & 4 \\
\enddata
\tablerefs{
(1) Millar et al. 1997; (2) Wishart 1979; (3) Glover \& Abel 2008; (4) Tilelens \& Hollenbach 1985; (5) Draine \& Bertoldi 1996; (6) Hollenbach \& McKee 1989; (7) Shapiro \& Kang 1987; (8) Nelson \& Langer 1997; (9) Lee et al. 1996
}
\end{deluxetable*}

\begin{deluxetable*}{lll} \label{t2}
\tablewidth{0pt}
\tablecaption{Cooling and Heating Processes}
\tablehead{Process &  Note  & Reference }
\startdata
Photoelectric heating by PAHs & depends on $G_0$ (Av=0 is assumed) &  1,\,2 \\
Cosmic ray heating & --- &  3 \\
H$_{2}$ photo-dissociation heating & depends on $G_0$ ($N_{\rm{H}_2}$ = Av=0 is assumed) & 4 \\
Ly-$\alpha$ cooling & --- & 5 \\
C$^{+}$ cooling (158$\mu$m) & escape probability =1 is assumed & 6 \\
O cooling (63$\mu$m) & escape probability =1 is assumed & 2,\,6 \\
CO ro-vibrational cooling & escape probability =1 is assumed & 7,\,8,\,9\\
Cooling due to rec. of  \\
electrons with grains \& PAHs & --- & 1 \\
H$_2$ cooling & collisions with H, H$_2$, He, p, \& e & 10 \\
\enddata
\tablerefs{
(1) Bakes \& Tielens 1994; (2) Wolfire et al. 2003; (3) Goldsmith \& Langer 1978; (4) Black \& Dalgano 1977; (5) Spitzer 1978; (6) de Jong et al. 1980; (7) Hollenbach \& MacKee 1979; (8) Hosokawa \& Inutsuka 2006; (9) Hollenbach \& McKee 1989; (10) Glover \& Abel 2008
}
\end{deluxetable*}

\subsection{Stability criterion}
According to the linear stability analysis for gas contracting isobarically with cooling (Koyama \& Inutsuka 2000; see, also Balbus 1995), 
thermal instability develops if the net cooling rate per unit volume $\Lambda$ increases with decreasing temperature:
\begin{equation}\label{crit}
\left( \frac{\partial \Lambda}{\partial T}\right)_{p}<0,
\end{equation}
where the subscript denotes the invariant quantity under the differentiation.
Note that the same criterion holds for the thermal instability both in thermal equilibrium states (Field 1965) 
and in an isochorically cooling medium (Schwarz et al. 1972; Burkert \& Lin 2000).
Since our background solution ensures the isobaric condition and that the temperature decreases monotonically with time, criterion (\ref{crit}) is equivalent to the condition that the net cooling rate increases with time:
\begin{equation}
\frac{d \Lambda}{d t}>0.
\label{critn}
\end{equation}

In a thermally unstable medium, density fluctuations with wavelength $\lambda$ grow exponentially with time.
The growth timescale is given by the cooling time $t_{\rm cool}\equiv \rho e/\Lambda$ for scales in the range
\begin{equation}
l_{\rm F} \lesssim \lambda \lesssim l_{\rm ac},
\label{eq:TIrange}
\end{equation}
where the lower and upper bounds are
the Field length (Field 1965)
\begin{equation}
l_{\rm F}= \sqrt{\kappa T/\Lambda},
\end{equation}
and the acoustic length
\begin{equation}
l_{\rm ac}= c_{\rm s}\,t_{\rm cool},
\end{equation}
respectively.
Here, $c_{\rm s}$ denotes the adiabatic sound speed, 
and $\kappa=2.5\times10^3\,T^{1/2}$ erg cm$^{-1}$ s$^{-1}$ K$^{-1}$ (Parker 1953) the thermal conductivity in the neutral atomic gas.
Fluctuations below the Field length ($\lambda \lesssim l_{\rm F}$) are damped by thermal conduction, 
while those above the acoustic length ($\lambda \gtrsim l_{\rm ac}$) cannot be causally connected within the cooling time 
and thus have a slower growth rate.
Although the most unstable scale is given by $\sqrt{l_{\rm F}\,l_{\rm ac}}$, all the perturbations in the range of 
equation (\ref{eq:TIrange}) grow roughly at a similar rate of $t_{\rm cool}^{-1}$ (Field 1965).

\begin{figure}[t]
\includegraphics[scale=0.6]{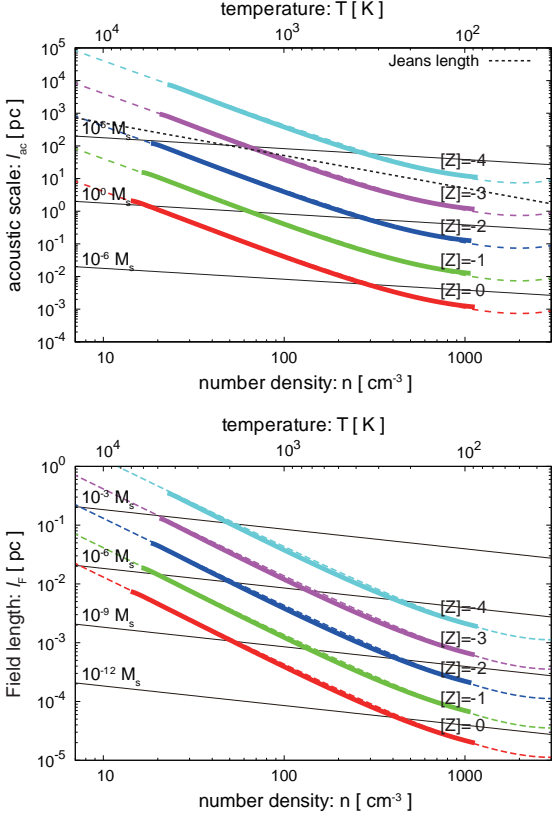}
\caption{\label{f1}
The acoustic scales (thick solid lines in the top panel) and the Field lengths (thick solid lines in the bottom panel) for the case of $G_0=1$ as a function of background density. 
The length of the lines indicates the density range over which the instability criterion is satisfied.
Upper horizontal scale shows corresponding temperature.
Red, green, blue, magenta, and aqua represents the results of $[Z]=0,\,-1,\,-2,\,-3,$ and $-4$, respectively.
Analytic lines of eq. (\ref{fitac}) for the acoustic scale and eq. (\ref{fitF}) for the Field length are plotted as a dashed lines.
Thin solid lines indicate the mass scale involved in the spherical region of the diameter $l$.
As a reference, the Jeans length (dashed) is shown in the top panel.
}\end{figure}

\begin{figure}[t]
\includegraphics[scale=0.6]{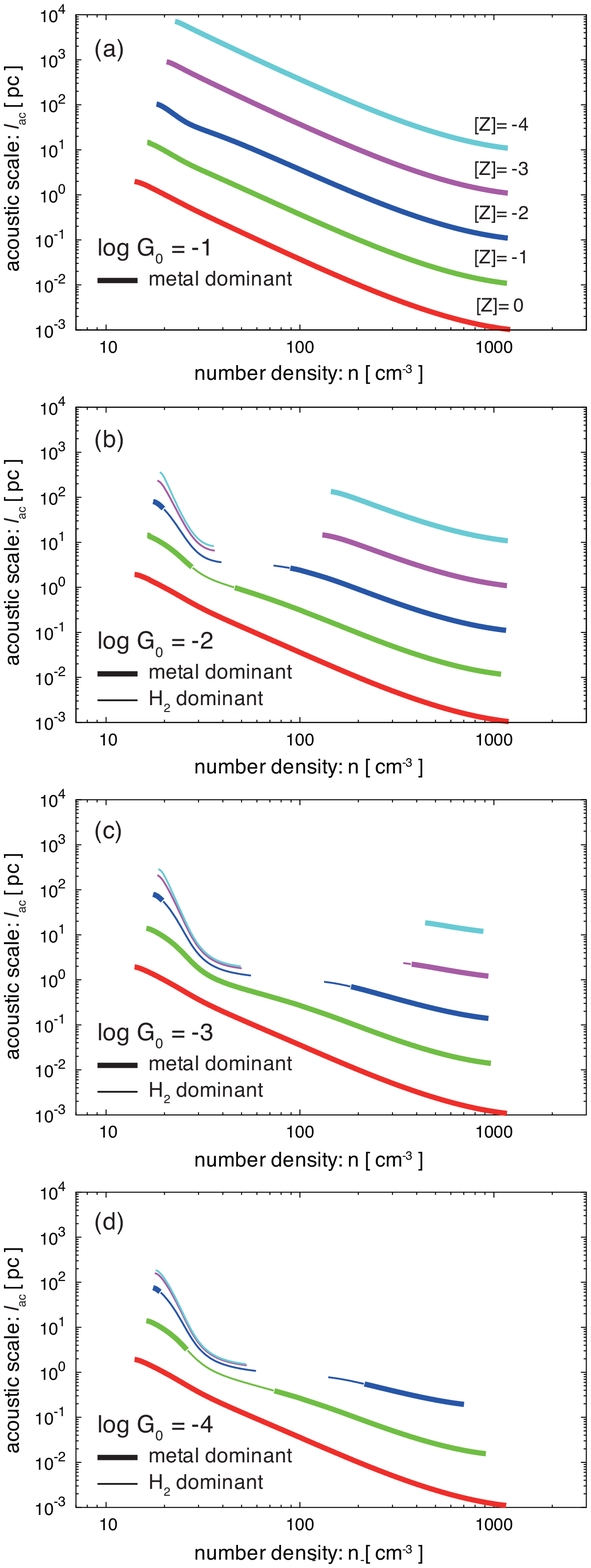}
\caption{\label{f2}
The acoustic scales for the cases of $G_0=10^{-1}$, $10^{-2}$, $10^{-3}$ and $10^{-4}$. 
The length of the lines indicates the density range over which the instability criterion is satisfied.
Red, green, blue, magenta, and aqua represents the results of $[Z]=0,\,-1,\,-2,\,-3,$ and $-4$, respectively.
Thick (thin) line is used when the dominant coolant is metals (H$_2$).
}\end{figure}

\subsection{Results of linear analysis}
As described in \S\ref{s11}, we numerically compute the isobaric contraction with $p/k_{\rm B}=10^5$ K cm$^{-3}$ from $(n,T)=(10 {\rm cm^{-3}}, 10000 {\rm K})$ to 
($2,000$ cm$^{-3}$, $50$ K) by integrating eq. (\ref{eq1}) and chemical reaction equations as the background model, and examine its thermal stability by using the criterion of eq. (\ref{critn}).
In Figure \ref{f1}, we show the acoustic (solid lines in the top panel) and Field lengths (solid lines in the bottom panel) 
as a function of background density for the case of $G_0=1$, i.e., the FUV strength is roughly the same as in the local ISM.
Red, green, blue, magenta, and aqua represent the results of $[Z]=0, -1, -2, -3$ and $-4$, respectively, and 
the solid line sections indicate the density range where the instability criterion is satisfied.
The thin solid lines indicate the mass scale involved in the spherical region of the diameter $l$.

The local ISM is known to be thermally unstable in the temperature range $100{\rm K} \lesssim T \lesssim 5000 {\rm K}$ 
owing to the fine-structure-line cooling by [CII] 158$\mu$m and [OI] 63$\mu$m lines (Wolfire et al. 1995, 2003; Koyama \& Inutuska 2000).
In the cases of $G_0=1$, since the cooling is always dominated by 
fine-structure lines, whose cooling rate can be fitted by (Koyama \& Inutsuka 2002)
\begin{equation}
\Lambda \simeq 3 \times 10^{-25}\,n^2\,T^{1/2}\,\exp (-T_{\rm line,2}/T)\,(Z/Z_{\sun})\,\mbox{erg cm}^{-3}\,\mbox{s}^{-1},
\end{equation}
the cooling time, the acoustic scale and the Field length can be written as:
\begin{eqnarray}
t_{\rm cool}&\simeq& 0.4\,\mbox{Myr}\,\,\left(Z/Z_{\sun}\right)^{-1}~n_1^{-3/2}~(p/k_{\rm B})_5^{1/2} \nonumber \\
&&\times \exp\left[10^{-2} T_{\rm line, 2}~n_1~(p/k_{\rm B})_5^{-1}\right],\\ \label{fitTC}
l_{\rm ac}&\simeq& 4\,\mbox{pc}\,\,\left(Z/Z_{\sun}\right)^{-1}~n_1^{-2}~(p/k_{\rm B})_5 \nonumber \\ \label{fitac}
&&\times \exp\left[10^{-2} T_{\rm line, 2}~n_1~(p/k_{\rm B})_5^{-1}\right],\\ \label{fitF}
l_{\rm F}&\simeq& 0.013\,\mbox{pc}\,\,\left(Z/Z_{\sun}\right)^{-1/2}~n_1^{-3/2}~(p/k_{\rm B})_5^{1/2} \nonumber \\
&&\times \exp\left[5 \times 10^{-3} T_{\rm line, 2}~n_1~(p/k_{\rm B})_5^{-1}\right],
\end{eqnarray}
where $n_1=n/1$~cm$^{-3}$, $(p/k_{\rm B})_5=(p/k_{\rm B})/10^5$~K~cm$^{-3}$, and $T_{{\rm line},2}=T_{\rm line}/100$~K.
These analytic lines are also plotted as dashed lines in the top and bottom panels, which agree well with the numerical results.

In the top panel of Figure \ref{f1}, we also plot the Jeans length $c_{\rm s}\sqrt{\pi/G\rho}$ as a dotted line.
Physically, when the acoustic scale is larger than the Jeans length, i.e., when the cooling timescale is longer than the free-fall timescale, the gas contract isothermally, shortening the cooling time more rapidly than the free-fall time, because $t_{\rm cool}/t_{\rm ff}|_{T}\propto n^{-1/2}$.
Such a medium begins to contract isobarically, once $t_{\rm cool}<t_{\rm ff}$ as a consequence of the isothermal contraction.
Thus, we can see from the top panel of Figure \ref{f1} that, although the media with $[Z]\lesssim -4$ is thermally unstable in principle, the instability cannot grow
since the acoustic length, i.e., the cooling time, becomes comparable to the Jeans length, i.e., the free-fall time, only for $n\sim1,000$ cm$^{-3}$ above which the gas is thermally stable.

In numerical simulations, typically 1/10 of the Jeans length is used as the refinment criterion for adaptive mesh refinement (AMR) in order to resolve the dynamics due to gravity (Truelove 1997; Heitsch 2001; Federrath et al. 2010).
We see that the Jeans criterion manages to capture the acoustic scale (the largest scale of the thermal instability) only 
for the $[Z] \lesssim -3$ cases.
Therefore, we should be careful about the use of AMR code when the Jeans criterion is applied.
In the next section, by the convergence test of nonlinear simulation, we show that we need to resolve the maximum scale of the thermal instability (the acoustic scale at $T\sim 5,000$ K) with more than 60 cells in order to follow the fragmentation.

For lower FUV strength $G_0$, thermally stable H$_2$ cooling can dominate the cooling.
Figure \ref{f2} shows the acoustic scales $l_{\rm ac}$ for different metallicities and with $G_0=10^{-1}$ (panel [a]), $10^{-2}$ (panel [b]), $10^{-3}$ (panel [c]), and $10^{-4}$ (panel [d]).
A thick line is used when the main coolant is metals, and a thin line is used when the main coolant is H$_2$.
The general tendency is that a weaker FUV field reduces the unstable domain by the metal cooling because of the H$_2$-cooling domination.
At low densities, unstable regimes by H$_2$ cooling appear.
If the H$_2$ fraction is constant, the H$_2$ cooling would not cause thermally instability.
In a dynamical medium, on the other hand, it can be unstable, because the H$_2$ fraction (or the amount of coolant) increases with density (i.e., the cooling rate can be enhanced as the temperature drops; $(\partial \Lambda/\partial T)_p<0$).

\begin{figure}[t]
\includegraphics[scale=0.6]{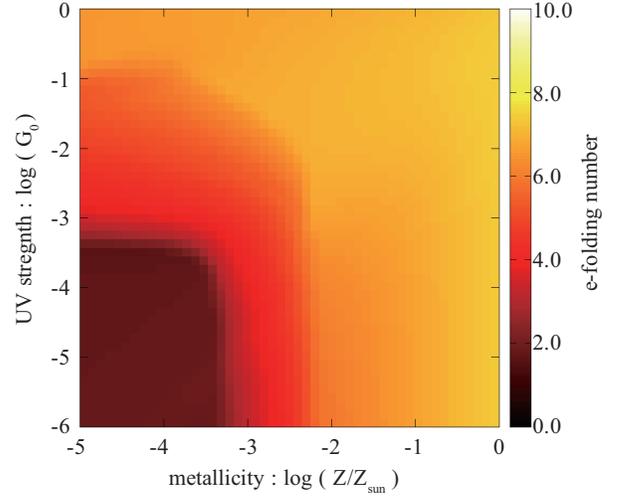}
\caption{\label{ff2}
e-folding number ($N_{\rm f}=\int_{d\Lambda/dt>0}\,t_{\rm cool}^{-1}(t)\,dt$) of the thermal instability in the isobarically contracting medium as a function of the metallicity and the FUV strength.
}\end{figure}

To indicate the strength of the thermal instability in a weak-FUV environment, we calculate the e-folding number of the thermal instability during background contraction from $n=10$ cm$^{-3}$ to 2,000 cm$^{-3}$:
\begin{equation}\label{ef}
N_{\rm f}=\int_{d\Lambda/dt>0}\,t_{\rm cool}^{-1}\,dt,
\end{equation}
where integration is done while the instability criterion eq.~(\ref{critn}) is satisfied.
Figure \ref{ff2} shows the e-folding number as a function of the metallicity and the FUV strength.
In the next section, based on nonlinear simulations, we show that $N_{\rm f}\gtrsim 3$ is necessary for the thermal instability to develop and create a bimodal density distribution, although the detailed threshold $N_{\rm f}$ would also depend on the amplitude of initial density fluctuation.

\section{Numerical simulation of post-shock flow}

\subsection{Method}
In this section, we examine the nonlinear growth of the thermal instability in 
a post-shock medium by means of three-dimensional hydrodynamics simulations incorporating with radiative cooling, heating, and chemical reactions.
The basic equations to solve are
\begin{equation}
\frac{\partial\,\rho}{\partial t}+{\bf \nabla}\cdot(\rho\,{\bf v})=0,
\end{equation}
\begin{equation}
\frac{\partial\,(\rho{\bf v})}{\partial t}+{\bf \nabla}\cdot(p+\rho\,{\bf v}\otimes{\bf v})=0,
\end{equation}
\begin{equation}
\frac{\partial \,(\rho\,e)}{\partial t}+{\bf \nabla}\cdot\{(\rho\,e+p)\,{\bf v}\}\!=\!{\bf \nabla}\cdot(\kappa{\bf \nabla} T)-\Lambda,
\end{equation}
along with the ideal-gas equation of state
\begin{equation}
p= \frac{\rho k_{\rm B} T}{\mu m_{\rm H}}.
\end{equation}
The microphysics considered, i.e., chemical reactions, cooling/heating and thermal conductivity, are the same as those used in the linear analysis in \S\ref{s11}.
We impose a temperature floor of 40 K, corresponding to the CMB temperature $T_{\rm CMB}$ at redshift $z=14$.
We use a second-order Godunov method (van Leer 1979) for solving the hydrodynamical equations and 
a second-order piecewise-exact-solution method for the chemical reactions (Inoue \& Inutsuka 2008, 2012).
The cooling, heating, and thermal conduction terms are usually solved by using a 2nd order explicit scheme.
When the local cooling time is shorter than the 10\% of the CFL timestep, the cooling and heating terms are solved implicitly using the bisection method.

\subsection{Settings}
We create a post-shock cooling medium by colliding two counter-streams at the $x=0$ plane in a 3D box of extension $L_{\rm box}/2\ge x$ (also for $y$ and $z$) $\ge -L_{\rm box}/2$.
The average initial number density is set to be $\langle n_1 \rangle=2.5$ cm$^{-3}$, with density fluctuations with a power law spectrum with the Kolmogorov spectral index ($P_{\rho,\rm 3D} \propto k^{-11/3}$), where the bracket means spatial averaging.
The initial amplitude of the density fluctuations is chosen as $\Delta n_1/\langle n_1\rangle=0.1$, where $\Delta n_1$ is the dispersion of the initial density field.
The initial pressure is set to be constant at $p_1/k_{\rm B}=4,000$ K cm$^{-3}$ so that the density fluctuations are the entropy mode, which can be thermally unstable.
The colliding counter-streams with velocity $v_1$ create a shocked slab with pressure $\langle p_{\rm 2}\rangle \simeq \langle \rho_1\rangle v_1^2+p_1$, which contracts isobarically by radiative cooling as long as the converging flows are maintained.
The value of $v_1$ is chosen so that the average value of postshock pressure becomes approximately the same as that used for the linear analysis in Sec. 2, i.e., $\langle p_{\rm 2}\rangle /k_{\rm B}=10^5$ K cm$^{-3}$ ($v_1=16$ km s$^{-1}$)
\footnote{If we fix the $v_1$ irrespective of the metallicity, one may think that the growth of thermal instability would be limited in low-Z cases due to the small thickness of the shocked slab created by the converging flows compared to the $l_{\rm ac}$.
However, because the thermal instability can always grow in the $y-$ and $z-$directions, the fixed $v_1$ does not limit the growth rate of the thermal instability.}.
Since the acoustic scale, the maximum scale for the thermal instability, is inversely proportional to the metallicity, we scale the box size of the simulation as $L_{\rm box}=4\,(Z/Z_{\sun})^{-1}$ pc.
We perform five runs with $[Z]=0,\,-1,\,-2,\,-3$, and $-4$ under fixed FUV strength of $G_0=1$, and three runs with $G_0=-2,\,-3,$ and $-4$ under fixed metallicity of $[Z]=-3$.
We impose periodic boundary conditions at the $y,\,z=\pm L_{\rm box}/2$ boundary planes.
At the $x=\pm L_{\rm box}/2$ planes, we set $v(t,\pm L_{\rm box}/2)=\mp v_1$, $p(t,\pm L_{\rm box}/2)=p_1$, and $\rho(t,\,x=\pm L_{\rm box}/2,\,y,\,z)=\rho(t=0,\,\mp L_{\rm box}/2\pm v_{\rm 1}\,t,y,z)$, indicating that there are continuous inflows with density fluctuations through these boundaries.
Although our box scale $L_{\rm box}$ can be extremely large in low metallicity cases, e.g., $[Z]=-3$ and $-4$, 
this does not mean that medium of such large spatial extent is necessary for the thermal instability to grow.
Recall that thermal instability is active at all the smaller scale down to the Field length (see eq. [\ref{fitF}]).

The numerical domain is divided into $1024^3$ cubic homogeneous cells, so the numerical resolution is $\Delta x= 3.9\times 10^{-3} (Z/Z_{\sun})^{-1}$ pc.
With this resolution, the acoustic scale is resolved in almost entire density range, while the Field length is resolved only in the low density regime (see Figure \ref{f1}).
Thus, we have to bear in mind that, although structures below the resolution scale are not expressed in our simulation, 
clumps could be present even with smaller scales down to the Field length (Koyama \& Inutsuka 2004; Aota et al. 2013).
However, despite an insufficient resolution for the Field length, 
properties of the thermally bistable medium, e.g.,  the mass function of cold clumps, and the power spectra of velocity and density, are known to 
converge on large scales (Hennebelle \& Audit 2007; see also Gazol et al. 2005; V\'azquez-Semadeni et al. 2006).
This is because most of the mass of the cold gas created by thermal instability is contained in large clumps that are formed by the growth of large scale fluctuations (typical mass of the clumps will be discussed in \S4.2).

\begin{figure*}[t]
\includegraphics[scale=0.25]{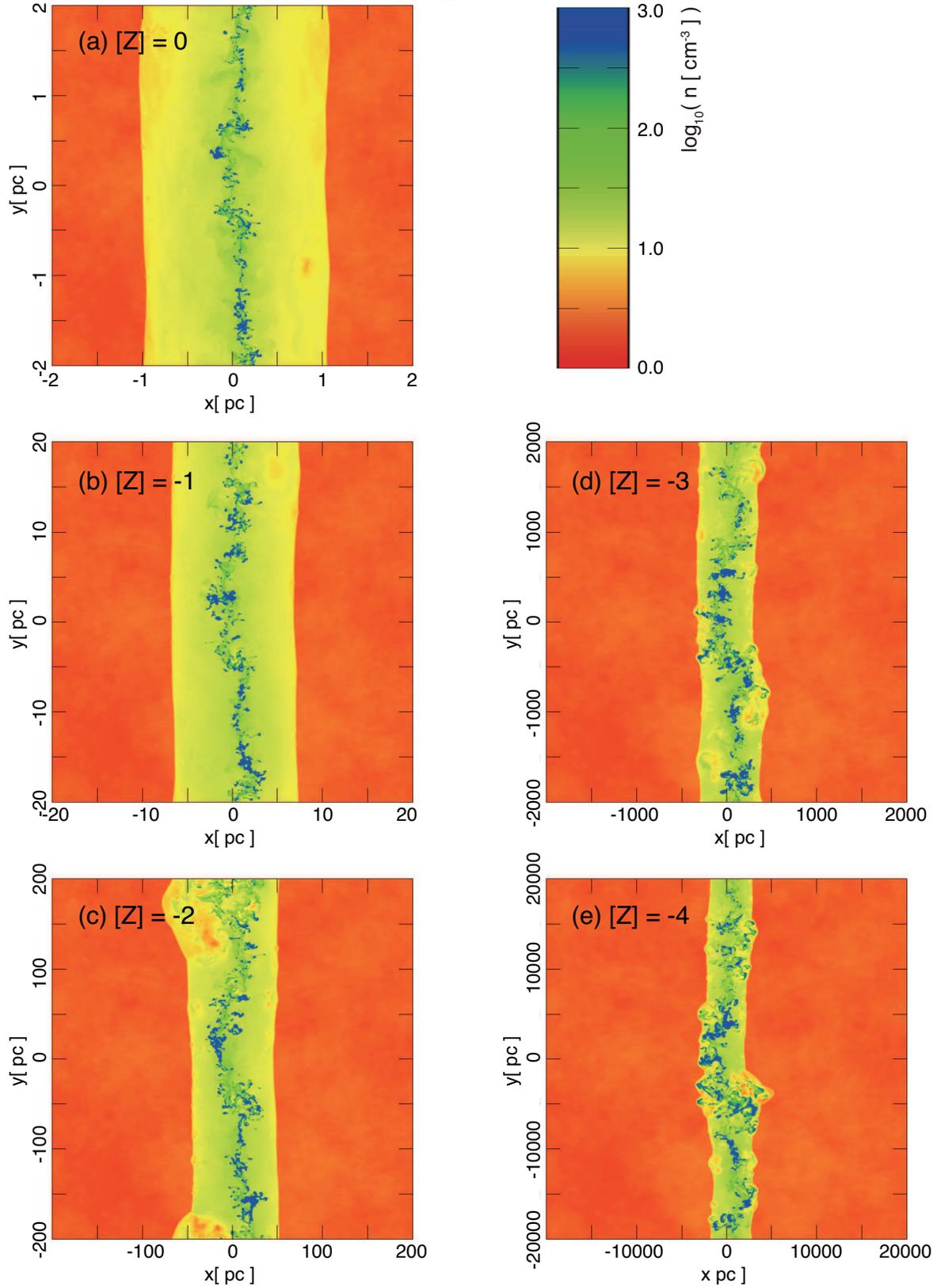}
\caption{\label{f3}
Density snapshots at the $z=0$ plane for the cases of $[Z]=0$ (panel a), $-1$ (b), $-2$ (c), $-3$ (d), and $-4$ (e) with $G_0=1$.
Data are sampled at $t=0.6\,(Z/Z_{\sun})^{-1}$ Myr, which corresponds to a few cooling time for the postshock gas.
}\end{figure*}

\begin{figure}[t]
\includegraphics[scale=0.65]{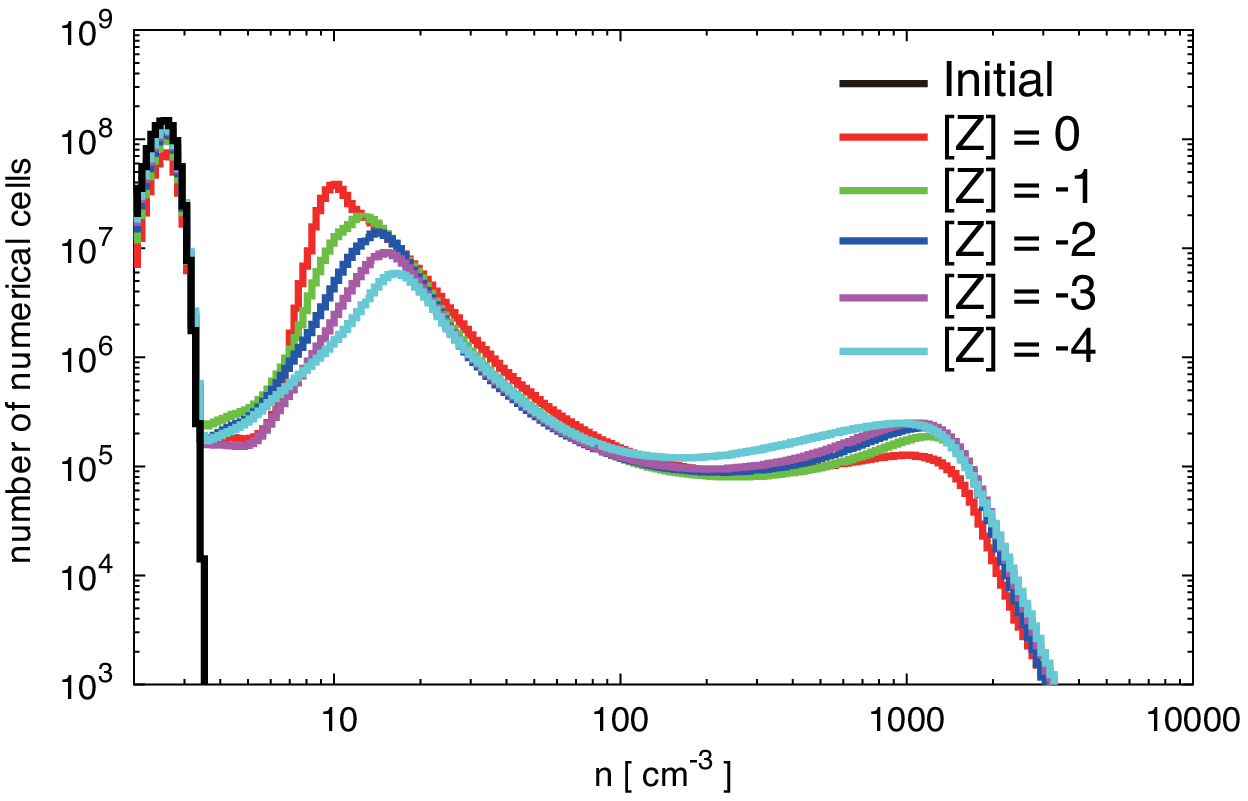}
\caption{\label{f4}
Probability distribution function of the density in the numerical domain for various metallicity models with $G_0=1$.
Red, green, blue, magenta, and aqua lines respectively represent the cases of $[Z]=0$, $-1$, $-2$, $-3$, and $-4$ at $t=0.6\,(Z/Z_{\sun})^{-1}$ Myr, and black line shows initial (preshock) state.
}\end{figure}

\subsection{Results}
\subsubsection{$G_0=1$ cases}
 In Figure \ref{f3}, we show snapshots of the density distribution on the $z=0$ plane for four different metallicities 
$[Z]=0$ (panel a), $-1$ (b), $-2$ (c), $-3$ (d), and $-4$ (e) with the same FUV strength $G_0=1$.
The snapshots are taken at $t=0.6\,(Z/Z_{\sun})^{-1}$ Myr, which corresponds to a few cooling times for the gas just behind the shock ($\langle n \rangle_{\rm ps}\simeq 10$ cm$^{-3}$, see eq. [\ref{fitTC}]).
In all the cases, clump formation by thermal instability is clearly visible in the shocked slabs.
Note that the simulation run-time above is required just because we start the simulations from such a low density as $n_{\rm ps}\simeq 10$ cm$^{-3}$.
In more realistic situations for the medium with $[Z]\gtrsim -4$, as we have discussed in \S 2.3, an initial low-density gas contracts 
rapidly until the cooling time becomes comparable to the free-fall time.
This contraction makes the initial gas density for the growth of thermal instability much higher than the present setting of $n_{\rm sh}\sim10$ cm$^{-3}$. 
Thus, the thermal instability can create clumps in a timescale shorter the present run-time because the cooling timescale decreases with the density as $0.6 (Z/Z_{\rm sun})^{-1} (n_{0}/10 \rm{cm}^{-3})^{-3/2}$ Myr.
For instance, in the case of $[Z] =-3$, the thermal instability begins to grow after $n > 70$ cm$^{-3}$ (see top panel of Figure \ref{f1}), where the cooling timescale is only 1/20 of that in the gas with $n = 10$ cm$^{-3}$ (see eq. [\ref{fitTC}]).
Although the later beginning of the instability growth leads to the smaller e-folding number (eq. [\ref{ef}]), we expect that the shorter dynamical timescale of the thermal instability due to the higher 
initial density would compensate it.

A remarkable structural characteristic of the thermally unstable ISM is dense clumps embedded in a diffuse medium 
(Pikel'Ner 1968; Field et al. 1969; Koyama \& Inutsuka 2002; see also, Audit \& Hennebelle 2005, 2010; Vazquez-Semadeni et al. 2006; Heitsch et al. 2005, 2006; Inoue \& Inutsuka 2008, 2009, 2012; Hennebelle et al. 2008; Banerjee et al. 2009).
Such a bi-phasic structure results from the fact that dense perturbations grow exponentially with time 
while the density in the surrounding diffuse medium increases more slowly by the overall contraction.
Figure \ref{f3} demonstrates that, even in the lowest-metallicity case studied, $[Z]=-4$, 
the ISM has a bi-phasic structure consistent with the linear analysis (see Figure \ref{f1}).
Note that although the ambient diffuse gas itself is also cooling and will eventually be incorporated into 
the dense clumps, such a bi-phasic structure is maintained as long as thermally unstable gases are continuously supplied 
from the shock front (see, e.g., Inoue \& Inutsuka 2012 for a recent simulation).

The bi-phasic nature can be observed more quantitatively in the probability distribution function (PDF) of the density shown in Figure \ref{f4}.
The red, green, blue, magenta, and aqua lines, respectively, represent the cases of $[Z]=0$, $-1$, $-2$, $-3$, and $-4$ at $t=0.6\,(Z/Z_{\sun})^{-1}$ Myr, 
and the black line shows the initial (pre-shock) state.
The bimodal distribution of the post-shock gas ($n>10$ cm$^{-3}$) reflects the bi-phasic structure.

Because the preshock medium contains seed density inhomogeneities, the Richtmyer-Meshkov instability (Richtmyer 1960) is induced by 
interactions between the density fluctuations and the shocks, making postshock medium turbulent.
Inoue et al. (2013) recently showed that the velocity dispersion of the post-shock turbulence agrees well with the growth velocity of the Richtmyer-Meshkov instability: $\Delta v \simeq (\Delta \rho/\langle\rho\rangle)/(1+\Delta \rho/\langle\rho\rangle)\,\langle v_{\rm sh} \rangle$.

\subsubsection{Lower $G_0$ cases}

\begin{figure}[t]
\includegraphics[scale=0.18]{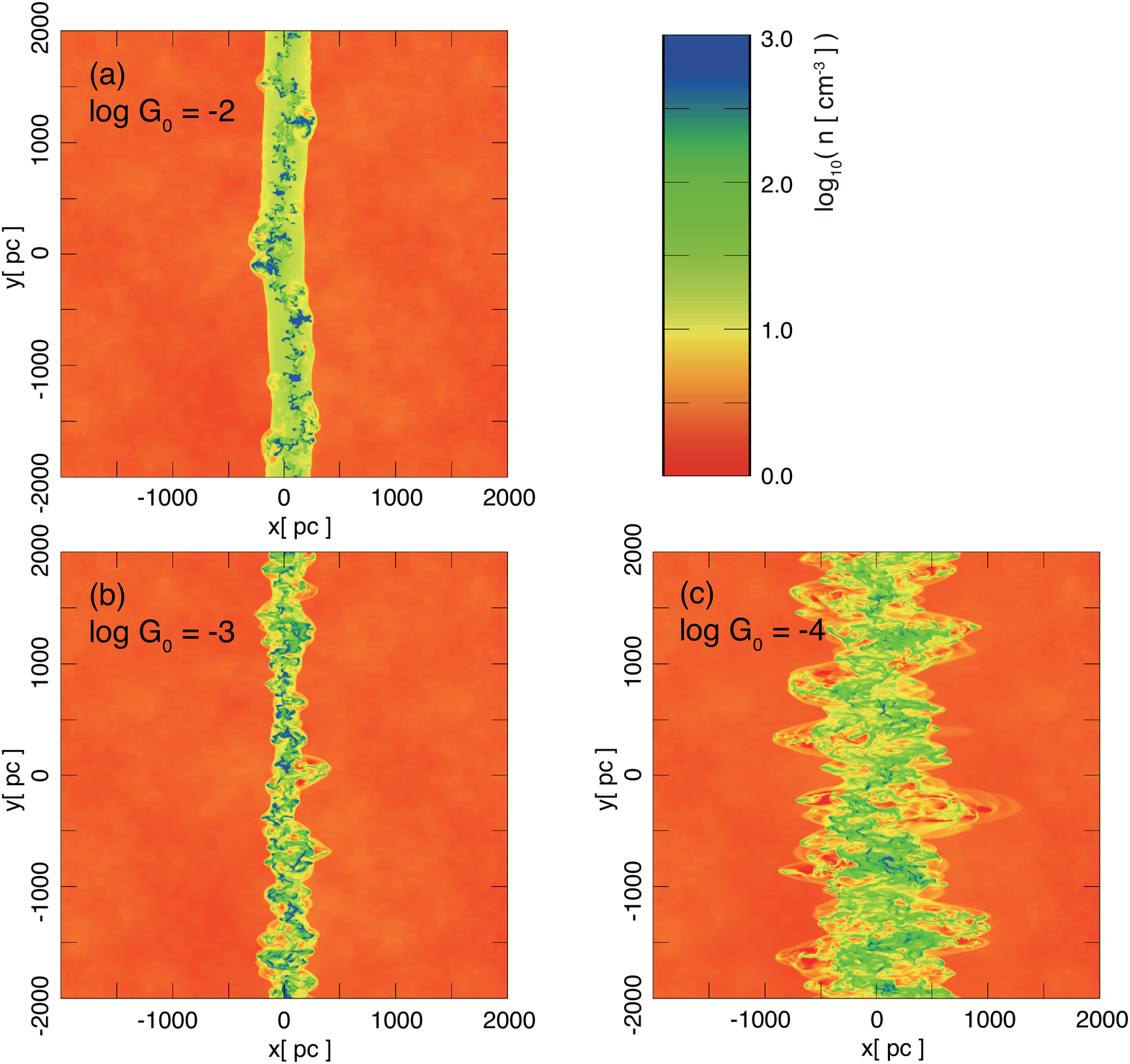}
\caption{\label{f5}
Snapshots of the density distribution at the $z=0$ plane for the cases of $G_0=10^{-2}$ (panel b), $10^{-3}$ (b), and $10^{-4}$ (c) with $[Z]=-3$
at $t=400$ Myr, corresponding a few cooling time for the $G_0\le10^{-3}$ cases.
}\end{figure}

\begin{figure}[t]
\includegraphics[scale=0.65]{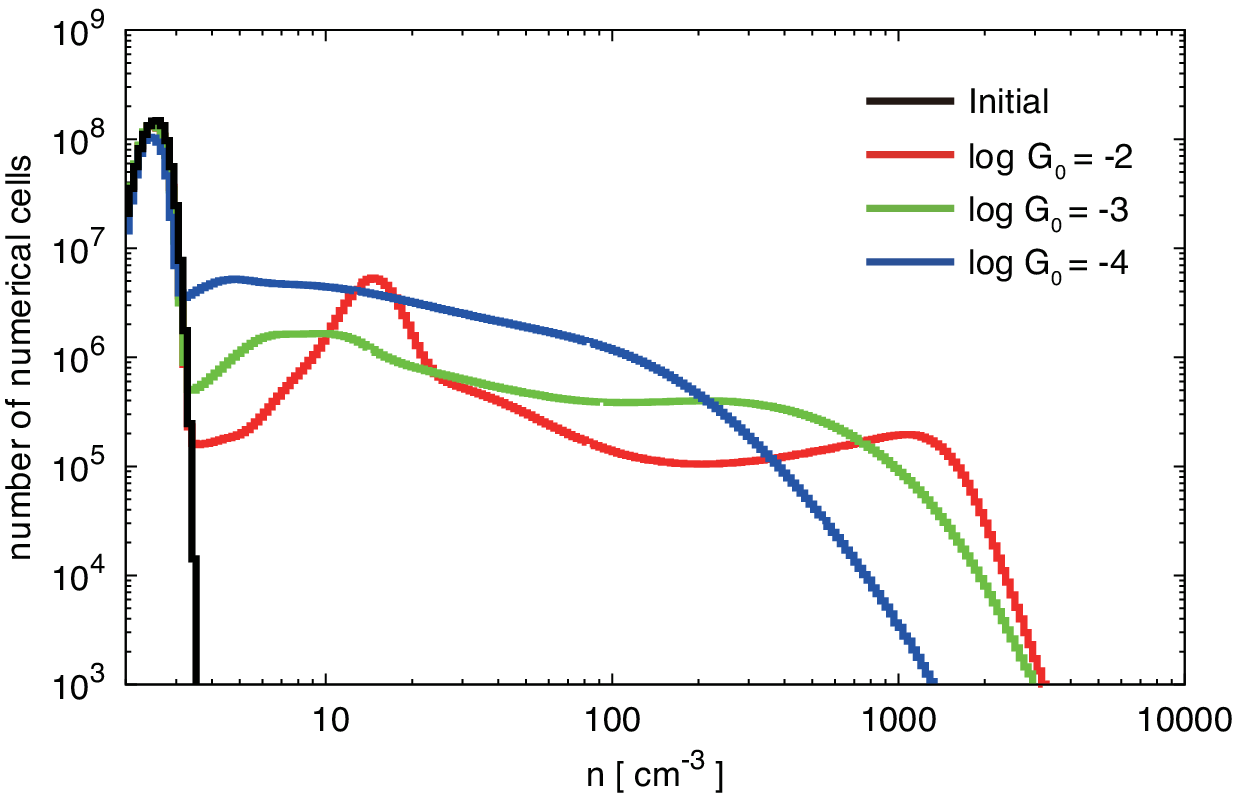}
\caption{\label{f6}
Probability distribution function of the density in the numerical domain for $[Z]=-3$ models with various $G_0$.
Red, green, and blue lines, respectively, represent the cases of $G_0=10^{-2}$, $10^{-3}$, and $10^{-4}$ at $t=400$ Myr, 
and black line shows initial (preshock) state.
}\end{figure}

To see the effect of enhanced H$_2$ cooling by $G_0$ reduction, we also perform simulations for $[Z]=-3$ models with $G_0=10^{-2}$, $10^{-3}$, and $10^{-4}$.
Their density snapshots and the density PDF are shown in Figures \ref{f5} and \ref{f6}, respectively.
We can see that the bimodal distribution becomes less remarkable as $G_0$ goes down and eventually disappears in the $G_0=10^{-4}$ case.
This is because a weak FUV field suppresses the effect of the thermal instability as shown 
by the linear stability analysis (middle panel of Figure \ref{f2}), which indicates that the thermally unstable density range shrinks when $G_0<10^{-2}$.
Recalling the e-folding number shown in Figure~\ref{ff2}, the vanishing of the bimodal distribution for $\log\,G_0\lesssim-3$ cases indicates that $N_{\rm f}\gtrsim3$ is necessary for the thermal instability to develop fully, although the detailed threshold $N_{\rm f}$ would also depend on the amplitude of the initial density fluctuation.
It should be noted that the contrasting result of $G_0=1$ run (effective thermal instability) and $G_0=10^{-4}$ run (ineffective thermal instability) 
is reminiscent of the difference of the two converging-flow simulations, one for the thermally bistable ISM and the other for (thermally stable) polytropic medium studied by Audit \& Hennebelle (2010).

Although not yet fully understood, the thermally biphasic medium could have some role in maintaining the interstellar turbulence.
Because the cold clumps are embedded in a higher temperature medium, shock dissipation is expected to be suppressed even if cold clumps move supersonically with respect to their internal sound speed.

\begin{figure*}[t]
\includegraphics[scale=0.65]{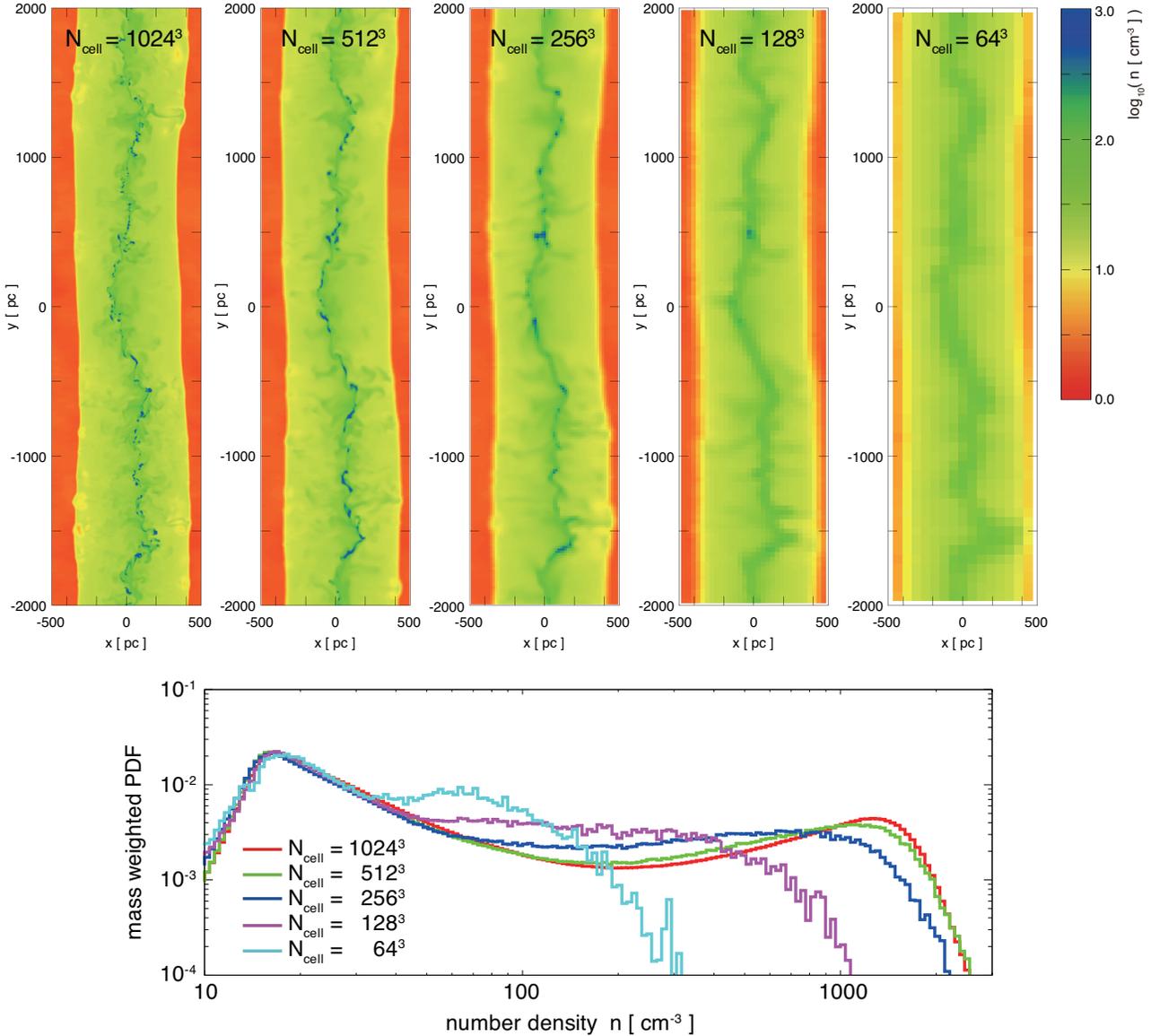}
\caption{\label{f7}
Upper panels: snapshots of the density distribution at the $z=0$ plane for the cases of $N_{\rm cell}=1024^3$, $512^3$, $256^3$, $128^3$, and $64^3$ with $[Z]=-3$ and $G_0=1$ at $t=220$ Myr at which the thermal instability begin to form dense clumps in the high-resolution runs.
Bottom panel: corresponding mass weighted PDFs.
}\end{figure*}

\subsection{The Minimum Resolution Required}
Here we consider how many numerical cells are necessary to capture the fastest growing mode of the thermal instability.
For this purpose, we perform additional simulations for $[Z]=-3$ and $G_0=1$ with different cell numbers of $N_{\rm cell}=1024^3$, $512^3$, $256^3$, $128^3$, and $64^3$.
In the top panels of Figure~\ref{f7}, we show the resulting density snapshots of these runs at $t=220$ Myr.
We also plot the mass weighted PDFs in the bottom panel.
At this time, dense clumps with $n\gg 10^3$ cm$^{-3}$ begin to appear for the case of the high-resolution runs with $N_{\rm cell}=1024^3,\,512^3$, and $256^3$.
However, in the cases of $N_{\rm cell}\le 128^3$, the dense fragments have not yet formed, indicating that even the fastest growing mode of the thermal instability is suppressed due to numerical diffusion (V\'azquez-Semadeni et al. 2003).

According to the results of the linear analysis in \S 2, the maximum scale of the thermal instability (the acoustic scale at $T\simeq 5000$ K) is given as $l_{\rm ac, max}\simeq 1\times (Z/Z_{\sun})^{-1}$ pc (eq [\ref{fitac}]), which is resolved by $N=l_{\rm ac, max}\,N_{\rm cell}^{1/3}/L_{\rm box}=0.25\,N_{\rm cell}^{1/3}$ cells.
Thus, the result of the above comparison indicates that, at least we need to resolve $l_{\rm ac, max}$ with more than 60 cells to follow the fastest condensation mode.
Note that, as we have mentioned in \S 3.2, the above minimum resolution does not ensure the convergence, since the convergence requires to resolve the Field length for capturing smallest fragments (Koyama \& Inutsuka 2004).
Even so, as we will discuss in \S 4.2, our criterion ($\Delta x \lesssim l_{\rm ac, max}/60$) would be useful, because large-scale clumps formed by the fastest growing mode have most of the mass of condensations for realistic seed fluctuations.

\section{Summary and Discussion}
\subsection{Findings}
In this paper, we have examined the linear stability and nonlinear growth of the thermal instability in isobarically contracting gas with various metallicities and FUV field strengths.
Our findings can be summarized as follows:
\begin{itemize}
\item When the H$_2$ cooling is suppressed by FUV fields ($\log G_0\gtrsim -3$), the interstellar medium (ISM) is thermally unstable
in the temperature range 100 K $\lesssim T\lesssim$ 7000 K owing to the cooling by 
[CII] 158 $\mu$m and [OI] 63 $\mu$m fine-structure lines regardless of metallicity.
In this case, the ISM becomes bi-phasic with a bimodal density distribution (Figure \ref{f4}).

\item The maximum (i.e., the acoustic scale $l_{\rm ac}$) and minimum scales (i.e., the Field length $l_{\rm F}$) of the thermal instability, 
which are well-fitted by eqs. (\ref{fitac}) and (\ref{fitF}), respectively, become smaller with increasing metallicity.
Comparisons of the nonlinear simulations with different resolution
indicates that the maximum scale of the thermal instability should be resolved more than 60 cells to follow the fastest growing mode of the thermal instability (\S3.4).

\item Under sufficiently weak FUV fields and with low metallicity, the density range of thermal instability shrinks owing to the dominance of H$_2$ cooling (Figure \ref{f2}).
As the FUV intensity is reduced, bi-phasic structure becomes less remarkable and disappears eventually (Figure \ref{f6}).
These results indicate that the effect of the thermal instability is suppressed when the e-folding number is $N_{\rm f}\lesssim3$ in the case of our initial conditions, which corresponds to the conditions $[Z]\lesssim -3$ and $\log G_0\lesssim -3$ (see, Figure~\ref{ff2}).
\end{itemize}

\subsection{Implications to the first galaxy formation}
According to the recent numerical studies on the first galaxy formation (Safranek-Shrader et al. 2014), the gravitational 
fragmentation scale is set by the Jeans mass when the temperature of the isobarically contracting gas reaches the CMB temperature floor.
The typical mass scale of such fragments is
\begin{eqnarray}
m_{\rm J,CMB}&\equiv &\frac{4\pi}{3}\rho\left(\frac{l_{\rm J}}{2}\right)^3 \nonumber \\
&=&190 \mbox{ M}_{\sun} \left(\frac{p/k_{\rm B}}{10^5\,\mbox{K cm}^{-3}}\right)^{-1/2}
\left(\frac{1+z}{10}\right)^{2}, \label{massG}
\end{eqnarray}
where $l_{\rm J}=c_{\rm s}\sqrt{\pi/G\rho}$ is the Jeans length and the CMB temperature written in terms of the redshift $T_{\rm CMB}=2.73\,(1+z)$ is used.
Safranek-Shrader et al. (2014) regarded this mass scale as the mass scale for stellar clusters in the first galaxy.
In their simulation, however, clump formation by thermal instability may not be resolved properly 
because of their employment of the Jeans criterion.
In the following, we speculate how the thermal-instability-induced fragmentation affects the fate of the ISM evolution.

The clump mass function by the thermal instability is known to take the power-law form:
\begin{equation}
N(m)\,dm\propto m^{(\delta-3)/3-2}\,dm, \label{eqHA}
\end{equation}
where $\delta$ is the power-law index of the three-dimensional power spectrum of the seed density fluctuations (Hennebelle \& Audit 2007; Hennebelle \& Chabrier 2008), which gives $N(m)\propto m^{-1.78}$ for realistic Kolmogorov-type fluctuations of $\delta=11/3$.
For this distribution, most of the mass is locked up in large fragments, 
with 50\% of the total mass being in the clumps of $m\gtrsim 0.04\,m_{\rm max}$.
Hence, we regard $m_{\rm TI}\equiv 0.04\,m_{\rm max}$ as the typical mass of the fragments due to the thermal instability.
Using the acoustic scale $l_{\rm ac}$ (eq. \ref{fitac}), the typical mass of the fragments can be written as
\begin{eqnarray}\label{massT}
m_{\rm TI}&\equiv& 0.04 \times \frac{4\pi}{3}\rho\left(\frac{l_{\rm ac}}{2}\right)^3 \nonumber \\
&\sim& 1.3 \times 10^4 \mbox{ M}_{\sun} \left(\frac{Z}{0.01Z_{\sun}}\right)^{-3}\left(\frac{T}{5000\,\mbox{K}}\right)^{5}\left(\frac{p/k_{\rm B}}{10^5\,\mbox{K cm}^{-3}}\right)^{-2},
\end{eqnarray}
where we have evaluated the mass scale at $T=5,000$ K at which $m_{\rm max}(T)$ takes the maximum.
Comparison of $m_{\rm TI}$ with $m_{\rm J,CMB}$ allows us to define the critical metallicity above which the typical clump mass falls below 
the gravitational fragmentation mass:
\begin{eqnarray}
Z_{\rm TI}&=&0.042\,Z_{\sun}\, \left(\frac{T}{5000\,\mbox{K}}\right)^{5/3}\left(\frac{p/k_{\rm B}}{10^5\,\mbox{K cm}^{-3}}\right)^{-1/2} 
\left(\frac{1+z}{10}\right)^{-2/3}.
\end{eqnarray}
We here discuss the cases when the thermal instability fully develops 
although the thermal instability condition depends also on the FUV intensity as seen in \S 2 and \S 3.
If the metallicity is much lower than the critical value ($Z\ll Z_{\rm TI}$), most of the clumps created by thermal instability are above the 
gravitational-fragmentation scale, 
suggesting that the thermal instability that happens prior to the Jeans instability could determine the mass scale of the stellar clusters in the first galaxy.
Fragmentation by thermal instability would lead to a power-law distribution of the cluster mass spectrum as given by eq. (\ref{eqHA}) with a high-mass cutoff $\lesssim m_{\rm max}$.
Note that the fact that the thermal instability grows only for $[Z]\gtrsim -4$ (see \S 2.3) indicates that the discussion 
above is applicable for $Z_{\rm TI}\lesssim Z\lesssim 10^{-4} Z_{\sun}$.
If $Z\gg Z_{\rm TI}$, on the other hand, most of clumps are much smaller than $m_{\rm J,CMB}$, which indicates that even when the gravitationally bound objects are formed by dense gas that reaches the CMB temperature, they would be in the form of complex of small-scale clumps, each of which is a gravitationally unbound object.
In addition, the clumpy medium formed by the thermal instability is highly turbulent which would substantially affect the star formation activity (see, e.g., Inoue \& Inutsuka 2012 for a recent simulation of such molecular cloud in the present-day ISM).
Thus, the star formation activity in the first galaxy with $Z\gg Z_{\rm TI}$ would be suppressed compared to those with $Z\ll Z_{\rm TI}$.
There could be a transition in star formation from a starburst mode to a more gradual one around this metallicity.

\acknowledgments
The numerical computations were carried out on XC30 system at the Center for Computational Astrophysics (CfCA) of National Astronomical Observatory of Japan and in part on K computer at the RIKEN Advanced Institute for Computational Science (No. hp120087).
This work is supported by Grant-in-aids from the Ministry of Education, Culture, Sports, Science, and Technology (MEXT) of Japan 
(23740154, 26287030 TI; 25287040 KO).

\end{document}